# Developer Investment in Agentic Platforms: Capability Spillovers, Hold-Up, and Portability

**Chupeng Xie**

## Abstract

Platforms increasingly host, distribute, and orchestrate third-party AI agents. A developer's investment in workflows, tool connections, testing, task records, and authorized memory creates a private capability stock and can also make complementary agents productive. The resulting stock is generated within a platform relationship, becomes platform specific, and is vulnerable to appropriation after investment. We model an origin platform that cannot commit to complete future fee and routing terms, a destination platform that commits to a verifiable recognition rule, and developers that invest before the origin resets terms. Portability protects investment through an exit option while preserving the origin's operating discretion. We characterize closed, partial, and full portability and show that the investment result survives incomplete ex post extraction. Verification has two opposing effects: it reduces transfer harm but can strengthen destination recognition and developer outside rent. It expands optimal portability only when safety and induced-investment gains exceed the additional rent and competitive leakage. Conditional on a target protected return, we compare portability with fee commitment and characterize the least-cost instrument mix. Finally, the origin is too closed when private leakage is mainly redistributive, but can be too open when duplication, fragmentation, or security costs are sufficiently real. The analysis turns agent interoperability into a conditional governance choice over relational capability rather than a presumption that openness is always efficient.

## 1. Introduction

An AI-agent platform does more than list software. It may host third-party agents, connect them to tools and data, assign tasks, coordinate multi-agent workflows, record performance, manage permissions, and collect payment. A procurement agent becomes useful after a developer builds supplier connectors, exception-handling routines, permission scopes, and a history of completed tasks. A contract agent becomes more valuable after it has been tested on a jurisdiction and integrated with document, identity, and approval systems. These investments create a capability stock that is valuable to the developer. They can also change what the platform as a whole can do. Adding a verified payment-and-identity agent, for example, may turn several agents that only recommend actions into a workflow that completes a purchase. The new agent is not merely another product attracting another user. It is a production input for other agents.

That productive role creates an investment-governance problem. The platform supplies execution infrastructure, distribution, identity, and customer access. The developer supplies workflows, tool integrations, testing, and task-specific knowledge. Once these investments are made, the platform may raise inference or tool fees, change revenue shares, reduce routing weight, alter certification requirements, restrict data access, or privilege a first-party agent. A developer anticipating such changes invests less. Yet much of the resulting loss falls on the platform: fewer tasks can be completed, complementary agents receive fewer calls, and the platform’s future fee base shrinks.

Why can the platform not simply promise to behave? Some future changes are efficient. Inference costs change, a tool becomes unsafe, an agent version fails, or demand shifts. A rule that freezes every future fee, routing weight, and access decision is incomplete or excessively rigid. The relevant governance instrument may instead be an exit right. If a developer can take an authorized and verifiable fraction of its workflow configuration, task record, reputation, and user-consented state to another compatible environment, the platform can still change its policy, but it cannot confiscate the entire return to sunk investment. Portability functions as a limited property right rather than as a complete long-term contract.

Portability is not free. It can help rivals free ride on investments made inside the incumbent ecosystem. A task record may be stale after the underlying model changes, valid for one task but misleading for another, or derived from data that the developer was never authorized to export. Workflow state can contain private information, tool credentials, and platform-specific dependencies. Unconditional migration may therefore create competitive leakage, privacy losses, counterfeit reputation, and unclear liability. The governance question is not whether openness is always desirable. It is how much effective portability a platform should credibly provide, what must be verified before a capability record travels, and whether surrendering lock-in can ever increase the incumbent's own profit.

We develop a sequential model with an origin platform, a potential destination platform, and third-party agent developers. Investment by developer $i$ creates a platform-specific capability stock $e_i$. It produces direct value and increases the expected value of workflows that use several agents. We first microfound this ecosystem term with tasks that require sets of agents: if a workflow succeeds only when each required component succeeds, investment in one agent raises the marginal productivity of its complements. The aggregate model then represents gross ecosystem revenue as

$$Y(e)=\sum_{i=1}^{N} F(e_i)+G\left(\sum_{i=1}^{N} e_i\right), \tag{1}$$

where $F$ is direct capability value and $G$ is the value of expanded task completion. This is a productive capability spillover, not simply a same-side or cross-side participation externality.

The origin cannot commit to its ex post access fee. In the sharp benchmark, once investment is sunk it leaves each developer only its portable outside return; a Nash-bargaining extension gives developers positive surplus and preserves the investment result whenever the origin retains bargaining power. An enforceable portability entitlement $0 \le p \le 1$ raises the recognized outside return and hence investment. The origin chooses $p$ before investment, internalizing that more portable developers build a larger fee base. Portability nevertheless requires the origin to leave more rent with developers and bears leakage and governance costs. The recognized-entitlement benchmark produces a marginal condition with a transparent decomposition:

$$N\Delta(e)e'(p)-Nr'(p)e(p)-C'(p)-T'(p)-(1-v)H'(p). \tag{2}$$

Here $\Delta(e)$ is the investment wedge: the marginal direct and ecosystem value that remains unprotected by the developer's post-investment return. Portability is profitable at the margin when the induced-investment term exceeds the added developer rent and governance costs. Section 6 decomposes the recognized return into the origin's entitlement and the destination's acceptance decision; the simple cross-partial is then replaced by an explicit conditional result.

Three results organize the paper. First, no commitment creates both a direct hold-up wedge and a productive-spillover wedge. The latter survives bargaining and even complete protection of direct value. Anticipating investment, the incumbent may voluntarily grant an exit right; endpoint marginal conditions partition the optimum into closed, partial, and full-portability regions rather than assuming that partial openness is best.

Second, portability requires a destination that recognizes the record. Better evidence raises recognition by reducing import harm, but it can also increase outside rent and leakage. Verification and portability are complements only when safety and induced investment dominate those rent channels. A comparison with fee commitment then shows why the instruments are non-nested: commitment restricts later policy, whereas portability preserves adaptation and protects exit.

Third, private and social portability have no universal ordering. The origin is too closed when developer rent and diverted business are mainly transfers, but the ordering reverses when migration creates sufficiently large duplication, fragmentation, or security cost. The resulting prescription is a scoped, recognized entitlement with evidence and liability, not unconditional export. A bounded software-policy audit remains in the Online Appendix and is not treated as evidence about developers or markets.

**Contributions**

The paper first contributes to platform openness and complementor investment (Boudreau 2010, Parker and Van Alstyne 2018). Those theories study access, control, and sequential innovation. We study an exit right after a developer has accumulated a platform-specific capability that the sponsor later taxes. Openness permits entry or building; portability protects the post-investment outside option.

The closest hold-up research uses governance or interoperability as commitment against later platform behavior. Bueno de Mesquita (forthcoming), for example, shows that governance can protect producer investment but impede efficient fee adjustment and upgrades. Our mechanism preserves later action and raises a destination-recognized exit value. We compare these non-nested instruments and show how productive cross-agent spillovers and recognition determine which is preferred.

Second, the paper connects incomplete contracting and property rights to a relational digital asset (Klein, Crawford, and Alchian 1978; Grossman and Hart 1986; Hart and Moore 1990; Williamson 1985). The residual right is neither platform ownership nor source-code ownership. It is a scoped right to move workflow state, performance evidence, authorized memory, and relationship metadata generated jointly by developer, platform, users, and interacting agents.

Data-portability models study user control, compliance, and competition (Cong and Matsushima 2026; Vijairaghavan, Hidaji, and Nault 2026). Our object is instead a productive capability record: an incumbent voluntarily chooses export rights to influence prior investment, and a destination decides whether to recognize the record.

Third, work on AI-agent markets asks how agents change transactions and market design (Hadfield and Koh 2025; Shahidi et al. 2025), while early experiments emphasize buyer choice (Allouah et al. 2025). We study the supply-side institution that determines whether third parties build capabilities that other agents use. Emerging communication and mandate standards make messages syntactically portable; our analysis asks which destination credits the accumulated evidence and which economic rights must accompany interoperability.

The rest of the paper proceeds as follows. Section 2 defines the institutional object and positions the paper. Section 3 introduces task composition and the sequential game. Sections 4–8 derive investment, portability, destination recognition, instrument choice, and welfare results. Section 9 provides numerical policy functions. Section 10 develops managerial and protocol implications. Section 11 concludes. Proofs, robustness extensions, the bounded software-policy audit, and a complete reproducibility map appear in the Online Appendix.

## 2. Institutional Setting and Related Literature

### 2.1 What is an agentic platform?

We use *agentic platform* for an organization that performs at least one of four functions for independently supplied AI agents: hosting execution, distributing agents to principals, routing tasks, or orchestrating multi-agent workflows. The definition includes enterprise agent platforms, agent marketplaces, professional-service networks, and environments that connect remote agents to tools and other agents. It does not require a consumer to browse a futuristic store of autonomous entities. The economic boundary is control over access to users, tasks, execution resources, or complementary agents.

The object in which a developer invests is observable before it is named. It includes domain workflows, API and data connectors, permission configurations, evaluation suites, exception handling, successful task traces, authorized memory, and established collaboration interfaces. These elements determine what an agent can do, under what scope, and how reliably it can be invoked. We call their economically usable stock *platform-specific agent capability*. “Specific” does not mean that every line of code is immobile. It means that the return to the bundle depends on platform-held state such as routing history, credentials, reputation, user relationships, and access to complementary agents.

The platform also accumulates an *ecosystem capability stock*: the expected value of the tasks its collection of agents can complete. This is not the platform’s model weight or compute stock. Let $Z$ denote possible workflows, $V_z$ the value of completing workflow $z$, and $P_z(e)$ its completion probability. Then

$$K(e) = \sum_{z \in Z} d_z V_z P_z(e) \tag{3}$$

is the value-weighted task set, with $d_z$ denoting task arrival. The distinction between private capability and ecosystem capability is central. A payment agent may earn little directly yet enable procurement, booking, and fulfillment agents to execute. Conversely, a popular stand-alone agent may add direct revenue without enabling other workflows.

Current protocols illustrate why capability and portability are separable. MCP standardizes resources, prompts, and model-controlled tools. A2A standardizes discovery through agent cards and task communication across opaque systems. AP2 adds verifiable credentials and mandates to bind authorization to payment. These layers can make a request readable across systems, but syntactic interoperability alone does not transfer a valid reputation, the user’s permission to move memory, or evidence that a new version retains an old version's capability. Economic portability requires both an entitlement and an evidence policy.

**2.2 Platform openness and complementor investment**

Digital platforms trade off external innovation against control. Boudreau (2010) distinguishes granting access from devolving control and shows empirically that alternative forms of openness have different innovation consequences. Parker and Van Alstyne (2018) model how platform openness and the duration of developer intellectual-property rights shape sequential innovation. Hagiu and Spulber (2013) examine first-party content as a coordination instrument in two-sided markets. Platform owners also selectively promote complements to manage ecosystem value (Rietveld, Schilling, and Bellavitis 2019), and empirical research documents how platform entry or core expansion changes complementor innovation (Foerderer et al. 2018; Wen and Zhu 2019).

Research on platform architecture and governance emphasizes that technical modularity, decision rights, and ecosystem relationships coevolve (Tiwana, Konsynski, and Bush 2010; Wareham, Fox, and Cano Giner 2014). Complementors create value jointly with an enterprise platform while making investments whose returns depend on platform access (Ceccagnoli et al. 2012). Theoretical work likewise distinguishes platform ownership and pricing in two-sided networks (Bakos and Katsamakas 2008). We retain that interdependence but isolate an exit-right decision that those frameworks do not solve: whether a record of accumulated capability remains economically usable after the sponsor changes policy.

Our platform also benefits from third-party investment, but the mechanism is not entry variety or the ability to build on an open core. Investment accumulates before the platform resets a fee. The governance variable determines how much accumulated capability can be used elsewhere after that reset. This timing yields a hold-up problem even if entry is open and APIs are public. Conversely, a closed execution environment can offer a strong portability entitlement by promising an authenticated export and accepting equivalent imports. Access openness, technical interoperability, and economic portability are three distinct margins.

The ecosystem term also differs from the classic anchor-tenant externality. Shopping malls subsidize anchors because they attract traffic to other stores (Pashigian and Gould 1998; Gould, Pashigian, and Prendergast 2005). An agent may likewise attract users, but our mechanism survives without additional traffic. Agent *i*'s investment raises output because another agent calls it inside a production chain. The externality is comparable to a reliability improvement at an upstream input that expands feasible downstream products. This productive complementarity is the main reason the AI-agent setting is more than a relabeled marketplace.

### 2.3 Hold-up, governance, and portability

Relationship-specific investment creates quasi-rents that can be appropriated after investment (Klein et al. 1978; Williamson 1985). Property-rights models allocate residual control to change investment incentives when complete contingent contracts are unavailable (Grossman and Hart 1986; Hart and Moore 1990). Platforms intensify this problem because a sponsor can change fees, ranking, access, and integration after many independent developers invest.

Fee commitment is one response, but it can be costly or incomplete. A permanent commission cap ignores changes in compute costs and risk. A fixed routing promise can prevent the removal of a failing agent. A prohibition on first-party entry can block efficient integration. Portability disciplines appropriation without specifying every future state: the platform may change terms, while the developer can exercise an outside option. This logic resembles an asset or termination right, but the portable object is generated within the relationship and must be scoped.

Verification is itself a costly control choice. Interfirm evidence rarely eliminates all residual risk; organizations balance the cost of controls against remaining performance and relational exposure (Anderson, Dekker, and Van den Abbeele 2017). Our verification variable has the same economic discipline but interacts with an endogenous property right. The platform values evidence more when it permits more state to move, and values the entitlement more when evidence makes movement safer. That two-way complementarity is absent when verification is modeled only as a fixed admission screen.

Data-portability research largely studies consumer switching, firm competition, recognition, privacy, and compliance (e.g., Krämer 2021; Cong and Matsushima 2026; Vijairaghavan et al. 2026). Vijairaghavan et al. analyze regulatory fines and investments that reduce firms' compliance cost; we analyze an incumbent's voluntary entitlement chosen to induce developer investment before hold-up. Our portable object is a productive capability stock, not a consumer data file. Reputation research shows that feedback can discipline behavior and that the design of identity and feedback systems matters (Dellarocas 2003; Cabral and Hortaçsu 2010). We add task and version applicability: an old success record is valuable only if it validly predicts performance under the destination's task and execution environment.

### 2.4 Scope

The baseline deliberately omits price competition between several symmetric origin platforms, self-preferencing, endogenous entry, user-side adoption, and private developer types. Each is important, but including all of them would obscure the investment-protection mechanism. The origin is not assumed benevolent; it chooses portability only when profitable. Section 6 endogenizes recognition by a destination platform, and Section 7 compares portability with fee commitment. Bargaining, heterogeneous spillovers, multihoming, and task-specific credentials are developed in the Online Appendix. Blockchain, trusted execution environments, zero-knowledge proofs, signed logs, and ordinary audits are treated as alternative ways to implement evidence. No architecture is assumed uniquely necessary.

# 3. Model

## 3.1 Task composition

There is an origin platform $A$, a potential destination platform $B$, and $N \geq 2$ third-party agent developers initially hosted by $A$. Developer $i$ chooses capability investment $e_i \geq 0$. For a workflow $z$, let $S_z \subseteq \{1, \ldots, N\}$ be the set of required agents and let $q_i(e_i) \in (0,1)$ be agent $i$'s success probability, with $q_i' > 0$ and $q_i'' \leq 0$. A serial workflow succeeds when each required component succeeds:

$$P_z(e) = \prod_{j \in S_z} q_j(e_j). \tag{4}$$

Substituting (4) into (3) gives a direct microfoundation for ecosystem capability. Define $Q_{z,-ij}$ as the success probability of all other components required by workflow $z$. If agents $i$ and $j$ appear in at least one common workflow, then

$$\frac{\partial^2 K}{\partial e_i \partial e_j} = \sum_{z: i, j \in S_z} d_z V_z q_i'(e_i) q_j'(e_j) Q_{z,-ij} \geq 0. \tag{5}$$

Thus investment in one agent raises the marginal product of its workflow complements. Equation (5) is the agent-specific primitive behind the aggregate spillover $G$. We use the symmetric reduced form (1), with $F' > 0$, $F'' \leq 0$, $G' > 0$, and $G'' \leq 0$. The direct term contains revenue attributable to the developer's own calls; $G$ contains task-completion value not contractibly attributed to a single input. Online Appendix A derives the reduced form from a large task system and shows that the main results also hold with heterogeneous $e_i$.

### 3.2 Portability and verification

At the time of investment, developer $i$'s portable outside operating return is

$$O(e_i, p) = r(p) e_i, r'(p) > 0, r''(p) \leq 0, \tag{6}$$

where $0 \leq p \leq 1$ is an export entitlement recognized by a reference destination. At $p = 0$, $r(0) = r_0 \geq 0$ captures code, general skill, or customers the developer can use without the origin's portable record. At $p = 1$, the origin honors the maximum feasible entitlement defined by the institutional environment; it need not imply that private user data, platform-owned tools, or third-party secrets move without permission. Sections 4–5 use the recognized-entitlement benchmark. Section 6 decomposes effectiveness into origin $A$'s entitlement $p$ and destination $B$'s recognition $q$, so that $r = r_0 + \mu p q$.

Verification reliability $0 \leq v \leq 1$ describes whether the ported evidence is authentic, authorized, current, and task applicable. It can reduce expected transfer harm at both platforms and make $B$ more willing to recognize an import. The distinction is institutional: $p$ allocates an export right, $q$ is the destination's acceptance decision, and $v$ governs the evidence on which that decision is made.

Portability creates three costs. $C(p)$ is a real resource cost of export, translation, and reconfiguration. $T(p)$ is competitive leakage borne by the incumbent; some or all of it is a transfer to users, developers, or rival platforms. $H(p)(1-v)$ is expected harm from unverifiable or misapplied claims. We assume $C', T', H' \geq 0$, convexity where needed, $H(0)=0$, and a verification cost $k(v)$, with $k'>0$, $k''>0$.

### 3.3 Timing and payoffs

The game has five stages.

1. Origin $A$ commits to export entitlement $p$ and evidence reliability $v$.
2. Destination $B$ observes $(p, v)$ and chooses how much of an imported record to recognize. Sections 4–5 normalize recognition to one; Section 6 solves this decision.
3. Developers simultaneously choose $e_i$, paying $c(e_i)$. The baseline uses $c(e)=c\,e^2/2$, $c>0$.
4. Investment is observed. Origin $A$ chooses access fees or equivalent routing/revenue-share terms. It cannot commit at Stage 1 to these complete state-contingent terms.
5. Each developer stays, migrates, or multihomes when permitted; tasks are produced.

The commitment asymmetry is institutional rather than behavioral. Destination recognition is a public rule over a finite credential schema: a developer and an origin can verify whether $B$ credits an imported record with the announced weight. The origin's future fee and routing policy instead depends on compute cost, congestion, safety incidents, demand, and agent quality, so a complete state-contingent promise is unavailable. Appendix D.3 relaxes the asymmetry. If $B$ also cannot commit, destination-side hold-up lowers investment further and creates a role for reciprocal recognition agreements; none of the baseline results relies on assuming that recognition is technologically costless.

The platform makes a take-it-or-leave-it ex post offer. Provided the inside surplus can cover $O$, the offer leaves the developer exactly $O(e_i, p)$. This stark assumption isolates hold-up. It can represent an access fee, a revenue-share adjustment, or routing that produces the same retained payoff. Developer $i$'s Stage 3 objective is therefore

$$U_i(e_i; p) = r(p) e_i - \frac{c}{2} e_i^2. \tag{7}$$

With symmetric investment $e$, platform profit is

$$\Pi(p, v) = N F(e) + G(N e) - N r(p) e - C(p) - T(p) - H(p)(1 - v) - k(v). \tag{8}$$

Assumption 1 requires that total in-platform value is sufficient to satisfy the outside options for the parameter region analyzed. Assumption 2 requires strict concavity of the platform's reduced profit in $p$; primitive sufficient conditions are given in Online Appendix B. Neither assumption imposes an interior solution.

## 4. Hold-Up and Capability Investment

Developer $i$'s investment is independent of other developers in the no-commitment benchmark because the platform appropriates the ecosystem return ex post. The unique symmetric choice is

$$e(p) = \frac{r(p)}{c}, e'(p) = \frac{r'(p)}{c} > 0. \tag{9}$$

The social planner, holding the institutional costs fixed, chooses symmetric investment to maximize $N F(e) + G(N e) - N c e^2 / 2$. Its first-order condition is

$$F'(e^{FB}) + G'(N e^{FB}) = c e^{FB}. \tag{10}$$

**Proposition 1. Hold-up and productive-spillover underinvestment**

For any fixed $p$, investment $e(p)$ in (9) is unique and increasing in effective portability. If

$$\Delta(e; p) \equiv F'(e) + G'(N e) - r(p) > 0, \tag{11}$$

then $e(p) < e^{FB}$. The investment wedge decomposes as

$$\Delta(e; p) = [F'(e) - r(p)] + G'(N e). \tag{12}$$

Even if portability or bargaining protects all marginal direct value so that $F'(e) = r(p)$, investment remains below first best whenever $G'(N e) > 0$. The ordering is not an artifact of complete ex post extraction. If the developer has Nash bargaining weight $\theta \in [0, 1)$, its operating return is

$$u^{op}(e, p) = \theta \left[ F(e) + \frac{G(N e)}{N} \right) + (1 - \theta) r(p) e, \tag{12A}$$

and equilibrium investment satisfies

$$(1 - \theta) r(p) + \theta [F'(e) + G'(N e)] - c e = 0. \tag{12B}$$

For every $\theta < 1$, investment is increasing in portability and remains below first best whenever $F'(e) + G'(N e) > r(p)$. The residual wedge is

$$F'(e) + G'(N e) - c e = (1 - \theta) [F'(e) + G'(N e) - r(p)]. \tag{12C}$$

The proof follows from strict convexity of cost and comparison of (9) with (10); Online Appendix B proves the bargaining extension and gives its investment derivative. The decomposition matters empirically. A platform can reduce the first wedge by protecting developer returns, using a fee commitment, or giving bargaining power. The second requires a contribution subsidy, shared task revenue, or platform-chosen portability that is valuable because the platform internalizes ecosystem output. Calling both effects “network effects” would conceal different remedies.

The proposition also clarifies why a zero-investment result in an extreme closed platform is not necessary. Developers may retain general code, human capital, and off-platform customers, so $r_0 > 0$ and $e(0) > 0$. Hold-up is measured by the investment gap, not by whether investment literally vanishes. As platform-specific task records and user relationships become more important relative to portable code, $r_0$ falls and the gap grows.

## 5. The Platform’s Portability Choice

Substituting (9) into (8) yields the platform’s reduced problem. Differentiating gives

$$\Pi_p(p, v) = N\,\Delta(e; p)\,e'(p) - N\,r'(p)\,e(p) - C'(p) - T'(p) - (1 - v)\,H'(p). \tag{13}$$

The first term is the value of investment induced by a stronger exit right. It includes direct capability value and ecosystem capability value, net of the return already protected. The second term is the additional operating return the platform must leave on the entire installed capability stock. The remaining terms are real conversion cost, competitive leakage, and verification-sensitive harm.

Define $M(p, v)$ as the right-hand side of (13). Under Assumption 2, it is strictly decreasing in $p$. This generates a complete institutional partition.

**Proposition 2. Closed, partial, and full portability regions**

Suppose the reduced platform profit is strictly concave for $0 \leq p \leq 1$. For fixed verification $v$, the unique profit-maximizing portability level satisfies:

1. $p^* = 0$ if and only if $M(0, v) \leq 0$;
2. $p^* \in (0, 1)$ if and only if $M(0, v) > 0 > M(1, v)$, in which case $M(p^*, v) = 0$;
3. $p^* = 1$ if and only if $M(1, v) \geq 0$.

Consequently, a small portability entitlement strictly raises platform profit if and only if

$$N \Delta(e(0); 0) e'(0) > N r'(0) e(0) + C'(0) + T'(0) + (1 - v) H'(0). \tag{14}$$

Proposition 2 is intentionally conditional rather than universal. Partial portability is not automatically optimal. A platform closes when the investment base responds weakly, when most capability value is already protected by $r_0$, or when transfer risk and leakage are high. It opens fully when the ecosystem value and investment response remain strong even near $p = 1$ and marginal governance costs remain low. Partial portability appears between these regions.

The platform-profit result is not a claim that portability itself creates revenue. The entitlement commits the platform to surrender some ex post appropriation. It becomes profitable because Stage 1 investment is elastic to the protected outside return and because the platform captures part of the enlarged ecosystem value. The platform chooses to tax a larger base at a lower effective confiscation rate.

Comparative statics follow directly from (13). Let a parameter $\beta$ raise $G'$, representing a larger value of completing composed workflows. If strict concavity is preserved, $p^*$ is nondecreasing in $\beta$. A tax-compliance agent that only substitutes for an incumbent stand-alone service may receive little portability protection. An identity, payment, or permission agent that unlocks many downstream workflows can rationally receive stronger protection, even from a profit-maximizing platform. Similarly, higher portability translation cost, more severe competitive leakage, or a lower investment response shifts the platform toward closure.

The result distinguishes selective subsidies from portable rights. A cash subsidy paid before investment can raise $e$, but if the platform later appropriates the resulting return, the developer anticipates that the subsidy finances an asset it cannot defend. A portable entitlement changes the continuation payoff. In Online Appendix C we let the platform choose both an up-front subsidy and $p$. Subsidies target developers with high measurable spillovers; portability protects dimensions of investment whose future return is difficult to contract. The instruments are complements when spillovers are measurable but future appropriation remains severe, and substitutes when a complete long-term transfer contract is enforceable.

## 6. Destination Recognition and Verification

Effective portability is not a file-copy operation. Origin $A$ can grant an export entitlement, but the entitlement protects investment only if destination $B$ recognizes the exported record. Recognition is costly because a genuine record can be stale, unauthorized, or inapplicable to the destination’s task, model, and tool environment. Verification affects both sides of this decision: it can reduce real transfer harm, yet it can also make exit more valuable and increase the rent that $A$ must leave with the developer.

Let $q \in [0,1]$ be the share of the entitlement that $B$ recognizes. A developer's outside return and investment are $r(p,q) = r_0 + \mu p q$ and $e(p,q) = r(p,q)/c$. Let $m(p,q) = p q e(p,q)$ denote recognized imported capability. Destination $B$ receives concave value $D(m)$, pays convex recognition cost $K_B(q)$, and bears expected import harm $(1-v) L(pq)$. Its reduced payoff is

$$\Pi^B(q; p, v) = D(m(p,q)) - K_B(q) - (1-v) L(pq). \tag{15}$$

At an interior optimum, $B$ chooses recognition according to

$$D'(m) m_q - K_B'(q) - (1-v) p L'(pq) = 0. \tag{16}$$

Because $\Pi^B_{qv} = p L'(pq) \geq 0$, a unique recognition choice $q^*(p,v)$ is nondecreasing in verification. This gives destination recognition an economic microfoundation rather than placing it inside an arbitrary outside-return function. Recognition can also be nondecreasing in $p$ when imported capability has increasing differences in entitlement and recognition; the primitive condition is stated in Online Appendix D.

Origin $A$ anticipates $q^*(p,v)$. Write $S(e) = N F(e) + G(Ne)$, $r(p,v) = r_0 + \mu p q^*(p,v)$, $e = r/c$, and collect conversion, leakage, transfer, and verification costs after the destination response in $Q(p,v)$. The origin's reduced profit is

$$\Pi^A(p,v) = S(e) - N r(p,v) e - Q(p,v). \tag{17}$$

Let $s'(e) = F'(e) + G'(Ne)$ and $\Delta = s'(e) - r$. The full cross-partial is

$$\Psi(p,v) \equiv \Pi^A_{pv} = N\left([s''(e) - c] e_p e_v + \Delta e_{pv} - r_{pv} e - r_p e_v\right) - Q_{pv}. \tag{18}$$

The term $-Q_{pv}$ is the safety channel: when reliable evidence reduces a larger expected harm under a stronger entitlement, it is positive. The term $\Delta e_{pv}$ is the induced-investment channel. The remaining terms capture diminishing ecosystem value and the extra rent created when verification makes the destination more willing to recognize exit.

**Proposition 3. Destination recognition and conditional verification**

Suppose $B$'s payoff is strictly concave in $q$ and $A$'s reduced payoff is concave in $(p, v)$. Then:

1. destination recognition $q^*(p, v)$ is nondecreasing in $v$;
2. portability and verification are strategic complements for origin $A$ if $\Psi(p, v) \geq 0$ and strategic substitutes if $\Psi(p, v) < 0$;
3. a sufficient condition for complementarity is

$$-Q_{pv} + N\Delta e_{pv} \geq N\left(\left[c - s''(e)\right]e_p e_v + r_{pv}e + r_p e_v\right). \tag{18A}$$

When recognition is fixed and verification only reduces harm, $e_v = r_v = 0$ and $Q_{pv} = -H'(p)$, so the familiar positive cross-partial is recovered as a special case. Once recognition is endogenous, monotonicity is no longer automatic. Verification can lower $A$'s optimal portability if it mainly raises outside rent or competitive leakage and does little to reduce real harm. This conditional result is the economic content of "verification makes portability profitable."

The two-platform structure also exposes a coordination problem. In a simultaneous standards-adoption stage, if $A$ must pay a fixed export-integration cost and $B$ must pay a fixed recognition cost, $p=q=0$ can be an equilibrium because neither investment has value without the other. A positive pair can nevertheless raise both platforms' joint surplus. Online Appendix D gives sufficient conditions for this low-portability/low-recognition trap. A common schema, reciprocal-recognition agreement, or standard-setting organization can be valuable by reducing the two fixed costs or by coordinating adoption; it does not need to mandate full portability.

**Corollary 1 (low-export/low-recognition coordination trap).** Suppose positive export and recognition require fixed costs $kappa_A, kappa_B > 0$, and each platform's incremental return is zero when the other chooses zero. Then $(p,q)=(0,0)$ is a Nash equilibrium. If some $(\hat{p},\hat{q}) \gg 0$ gives both platforms positive incremental payoff net of their fixed and variable costs, the zero equilibrium is Pareto dominated. A reciprocal-recognition agreement that conditions each investment on the other's adoption can implement the positive pair without requiring $p=q=1$.

The corollary explains why a technically available interchange schema need not create economic portability. Each side can rationally wait for the other to incur the complementary fixed cost. The governance problem is coordination over entitlement and recognition, not a mechanical preference for maximum openness.

### 6.1 What evidence should travel?

A platform should not export a scalar "agent score" detached from context. Let a portable credential be indexed by task $z$, agent version $m$, evidence procedure $a$, authorization scope $s$, and time $t$:

$$R_{i,z,m,a,s,t} = (\text{claim}, \text{issuer}, \text{task}, \text{version}, \text{procedure}, \text{scope}, \text{time}). \tag{19}$$

A destination can then decide whether the tuple is applicable. A universal rating collapses task difficulty, model changes, tool access, authorization, and evidence quality. Task- and version-bound evidence may cost more to produce, yet it can reduce $L$ and $Q$ enough to support greater recognition and a stronger entitlement. Online Appendix L maps this principle to current protocol layers and shows why identity, mandate, execution, and outcome evidence must remain separable.

## 7. Portability versus Fee Commitment

An enforceable fee or routing commitment also protects investment. Let $x \geq 0$ be the increment to the developer's retained marginal return produced by such a commitment. With destination recognition $q$, total protected marginal return and investment are

$$R(x, p; q) = r_0 + x + \mu p q, \quad e(x, p; q) = \frac{R(x, p; q)}{c}. \tag{20}$$

For investment incentives, $x$ and recognized portability $pq$ are substitutes. Institutionally, they are different. Commitment restricts the origin's future fees or routing and therefore bears enforcement and adaptation cost $J(x)$. Portability preserves that discretion but bears export, leakage, and transfer cost $Q(p, v; q)$. Conditional on a target protected return $z = x + \mu p q$, the platform solves

$$C(z; q, v) = \min_{x, p \geq 0} \{J(x) + Q(p, v; q) : x + \mu p q = z, 0 \leq p \leq 1\}. \tag{21}$$

**Proposition 4. The optimal protection instrument**

Suppose $J$ and $Q$ are increasing and strictly convex, $q > 0$, and a target $z$ is feasible. Then the cost-minimizing protection mix is unique. At an interior solution,

$$J'\left(x^{*}\right)=\frac{Q_{p}\left(p^{*},v;q\right)}{\mu q}. \tag{22}$$

All protection is supplied by portability if its marginal institutional cost per unit of protected return is below $J'$ over the feasible allocation; all protection is supplied by commitment under the reverse inequality; otherwise the platform mixes the instruments according to (22). For a target $z \le \mu q$, portability strictly dominates a pure fee commitment if and only if

$$Q\left(\frac{z}{\mu q},v;q\right)<J\left(z\right). \tag{22A}$$

Proposition 4 provides the non-nested comparison with platform-governance commitment. A veto, commission cap, or long-term routing rule can protect investment by restricting later action. A portability entitlement protects the same target return through a destination-recognized exit option and leaves the origin free to adapt. Higher contract incompleteness, compute-cost volatility, or safety-driven routing changes raise $J$ and shift protection toward portability. Higher privacy, staleness, or competitive-import cost raises $Q$ and shifts it toward commitment. Better destination recognition increases the amount of protected return produced by one unit of entitlement, but it may also increase leakage inside $Q$; both effects appear in (22).

Developers differ in capability spillovers. An agent that completes a narrow stand-alone task may have high direct value but low ecosystem contribution, while an identity, permissions, payment, or evaluation agent may appear in many workflows. A common export schema lowers translation and verification cost, while contribution payments target heterogeneous spillovers. Online Appendix C solves this separation and shows why discretionary developer-specific exit rights can reintroduce hold-up.

## 8. Private and Social Portability

Platform profit is not welfare. To isolate incidence, return to the recognized-entitlement benchmark; with endogenous destination recognition, the same expressions use total derivatives after substituting $q^*(p,v)$. Let $\omega \in [0,1)$ be the share of competitive leakage $T(p)$ that is a real resource cost rather than redistribution, and let $D(p)$ collect additional duplicated-infrastructure, fragmentation, or cross-platform security cost that the origin does not internalize. Given induced investment, welfare is

$$W(p,v) = N F(e) + G(N e) - \frac{N c}{2} e^2 - C(p) - H(p)(1-v) - k(v) - \omega T(p) - D(p). \tag{23}$$

The developer's operating return is a transfer. Differentiating at the same investment and verification level and using $c e = r(p)$ gives

$$W_p(p,v) - \Pi_p(p,v) = N r'(p) e(p) + (1-\omega) T'(p) - D'(p) \equiv \Omega(p). \tag{24}$$

**Proposition 5. Private closure, social closure, and reversal**

Suppose platform profit and welfare are concave in $p$. Then:

1. if $\Omega(p) \geq 0$ for every $p \in [0,1)$, the origin is weakly too closed: $p^P(v) \leq p^S(v)$;
2. if $\Omega(p) \leq 0$ for every $p \in [0,1)$, the origin is weakly too open: $p^P(v) \geq p^S(v)$;
3. if $\Omega$ changes sign, there is no global ordering without additional curvature restrictions.

The inequalities are strict for interior solutions when the corresponding wedge is strict. The familiar underprovision result is therefore a parameter region, not a universal policy claim. It applies when developer rent and diverted business are mainly redistributive. The ordering can reverse when portability causes substantial duplication, fragments scale economies, or creates security exposure that the origin does not bear.

Three interventions follow. First, a minimum export entitlement can address private closure, but only when $\Omega$ and the platform's marginal investment value justify it. Second, shared verification infrastructure can lower both platforms' evidence cost and reduce real harm without prescribing a corner entitlement. Third, contribution payments tied to composed-workflow value address the spillover component of (12), which portability alone need not eliminate. An undifferentiated data dump can raise $D$ and worsen welfare even when a scoped, recognized capability record would be beneficial.

## 9. Numerical Policy Functions

The analytical results characterize regions but do not show their size. We therefore evaluate a transparent functional form, not a calibration to a particular commercial platform. Let

$$F(e) = ae - \frac{b}{2}e^2, G(Ne) = \beta \log(1 + Ne), r(p) = r_0 + \mu p, \tag{25}$$

and let leakage and unverifiable-transfer harm be quadratic. The baseline parameters are $N = 6$, $a = 1.5$, $b = 0.8$, $c = 4$, $r_0 = 0.8$, $\mu = 2$, $T(p) = 0.4p^2$, $H(p) = 1.5p^2$, and $k(v) = 0.4v^2$. The script evaluates the complete $[0,1]$ choice set, verifies the analytical marginal-profit expression against central differences, and exports every policy point. The maximum absolute derivative discrepancy is $1.21 \times 10^{-10}$.

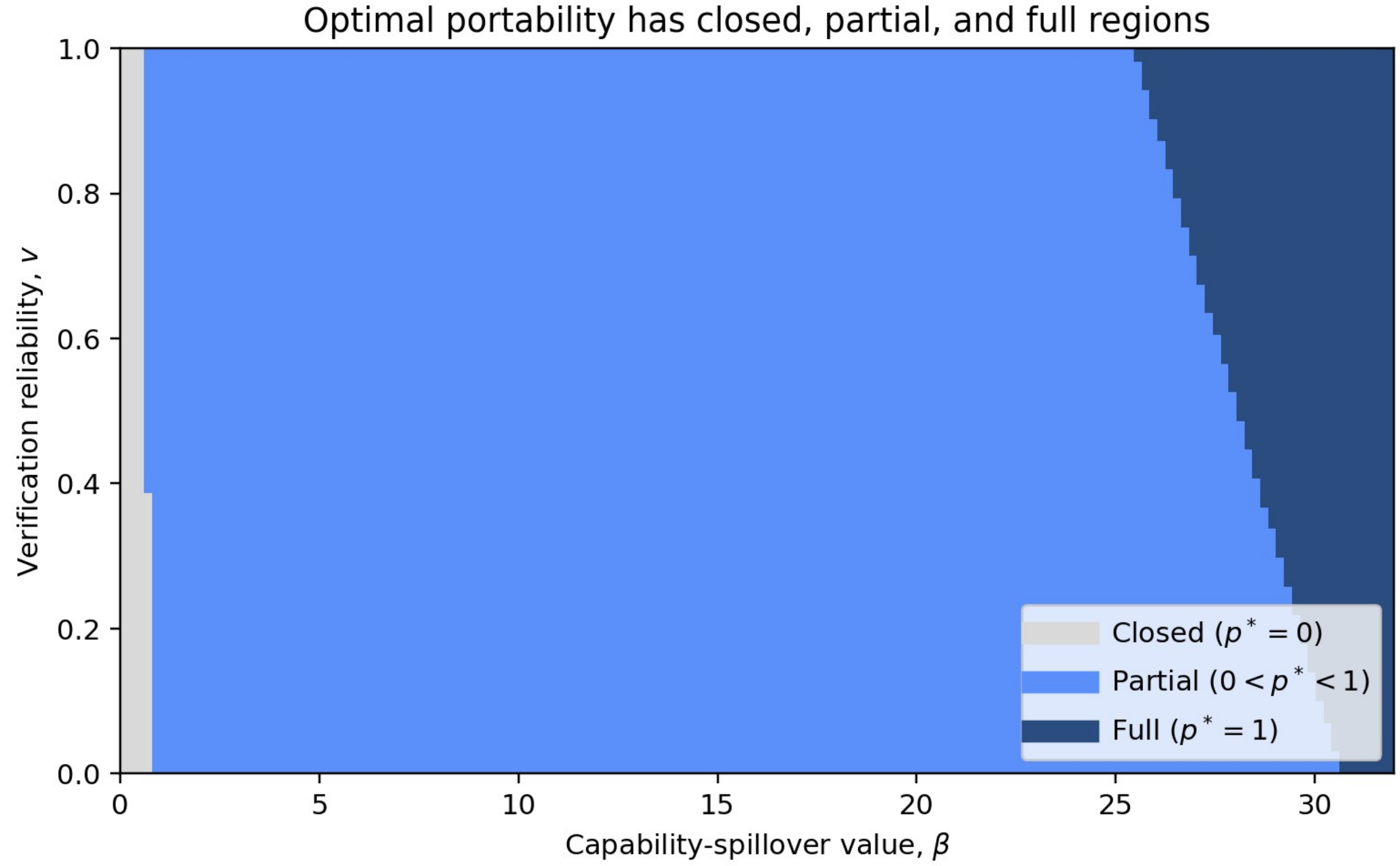

*Figure 1. Optimal portability has closed, partial, and full regions.*

Figure 1 varies ecosystem spillover value $\beta$ and verification reliability $v$ in the safety-channel benchmark with fixed destination recognition. All three regimes occur. At low $\beta$, induced investment cannot repay additional developer rent, so the platform closes even when verification is reliable. At intermediate $\beta$, the platform grants partial portability. At high $\beta$, the investment base remains valuable enough that the upper constraint binds. Increasing $v$ shifts both regime boundaries toward greater portability in this benchmark. Proposition 3 states the additional condition required when recognition and outside rent also respond to verification.

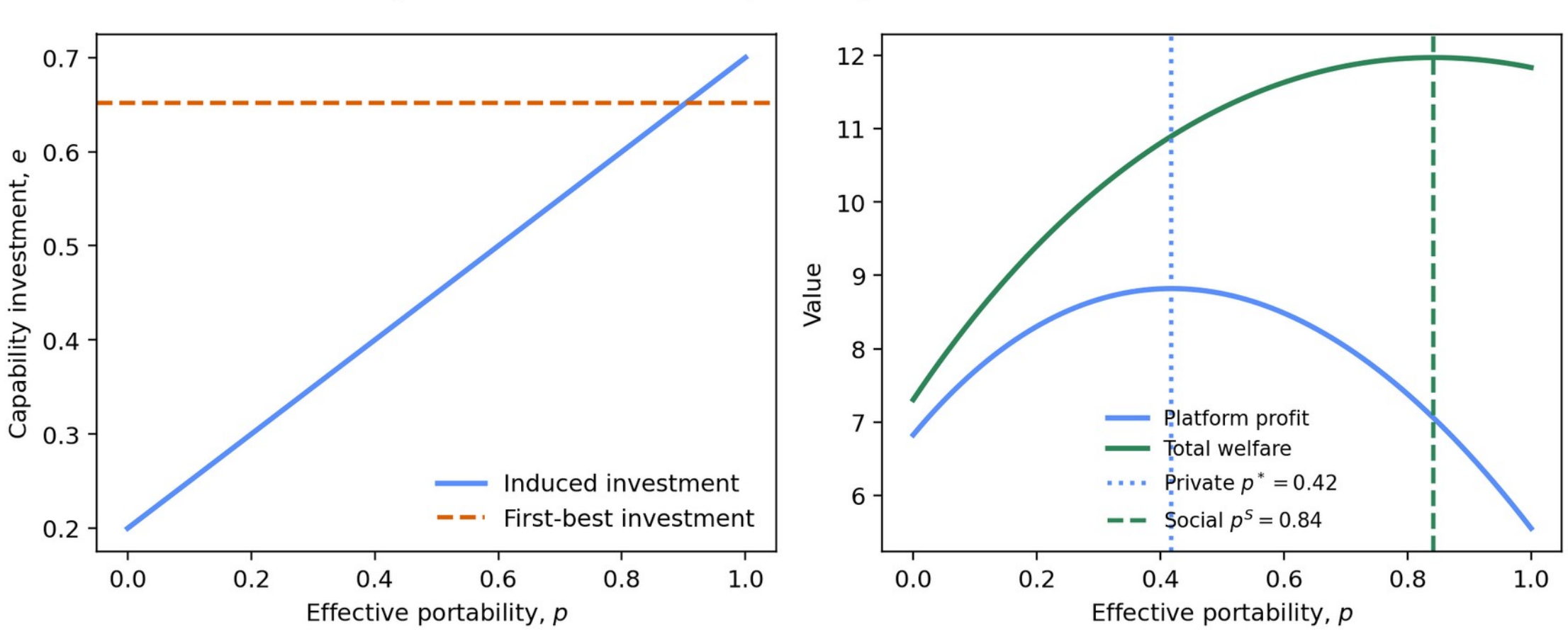

*Figure 3. Induced investment and the private–social portability wedge.*

Figure 3 separates investment and institutional choice under the redistributive-leakage benchmark $\omega = 0$ and $D = 0$. Recognized portability raises developer investment, but investment remains below first best over much of the range. Platform profit peaks before welfare because the platform subtracts the developer's improved outside return and competitive leakage. Under the baseline and $v = 0.75$, the private optimum is $p^P = 0.42$, while the social optimum is higher. The exact values are illustrative; Proposition 5 shows when the ordering holds and when real duplication or security cost can reverse it.

Online Appendix H reports the omitted profit curves and verification-comparative-statics witnesses. They show, respectively, the movement from closed to partial to full portability as spillover value rises, and both nonempty sign regions in Proposition 3 when destination recognition is endogenous.

# 10. Managerial and Protocol Implications

## 10.1 Price capability contribution, not just usage

Calls, users, and direct revenue are incomplete contribution measures. A platform should estimate how an agent changes the success probability and value of composed workflows. The relevant accounting identity is

$$\text{agent contribution} = \text{direct value} + \text{workflow spillover} - \text{compute and verification c} \tag{26}$$

An agent with modest direct volume can merit lower fees, verification subsidies, or development support if it unlocks many workflows. Conversely, a high-volume agent that consumes compute, duplicates existing capability, or creates correlated failure risk may contribute less than its transaction count suggests. Platforms already control task routing; the model implies that routing and developer programs should incorporate marginal ecosystem completion value rather than engagement alone.

The task-composition graph, rather than the agent's own call count, is the natural measurement unit. Online Appendix K develops randomized-routing, staged-release, and structural designs for estimating effects on complementary workflows.

## 10.2 Separate access, interoperability, and scoped portability

Managers often use "open" for three distinct choices. Access openness determines who may join. Technical interoperability determines whether systems can exchange valid messages. Economic portability determines whether an accumulated capability record retains value after exit. A public API can coexist with severe hold-up if routing history and credentials cannot move; a proprietary execution environment can still offer meaningful exit through authenticated export.

The useful artifact is therefore a scoped capability passport, not a universal score. It binds identity to task, model and tool versions, authorization scope, evaluation procedure, issuer, expiration, revocation, and export rights. A portability policy must state which fields travel, who can authorize them, and what a destination commits to recognize. The record may use a signed database, transparency log, trusted-execution receipt, or distributed ledger; the model ranks their economic cost and evidence properties, not their labels.

### 10.3 Preserve discretion while correcting the governance wedge

Platforms may resist commitment because they need to remove unsafe agents, change prices with compute cost, and update routing as quality changes. The analysis shows that portability can preserve this flexibility. The platform does not promise never to change a fee. It promises that specified, authorized evidence and state remain usable if it does. This converts an impossible complete contract into a bounded exit right.

The appropriate instrument depends on the asset. A short fee guarantee may fit an investment with a known payback period; portable task evidence fits long-lived capability whose future use is hard to specify; user memory may remain nonportable without consent. Private choice is not automatically efficient: the platform treats developer rent and rival gains as losses, while mandatory portability can create real duplication, fragmentation, or security cost. Proposition 5 identifies both orderings. Consortia and regulators should therefore target effective recognition, authorization, and shared verification rather than raw export or a universal maximum-entitlement rule.

## 11. Conclusion

Third-party agents can become productive infrastructure for an entire platform. Their developers invest in workflows, tools, testing, records, and relationships that create private capability and expand the platform's feasible task set. Because much of that capability becomes platform specific, a platform that cannot commit to future fees can appropriate its return and depress investment.

Portability changes this dynamic. It limits ex post appropriation through an exit option while preserving the origin's ability to update policy. The optimal entitlement can be zero, partial, or full. A profit-maximizing origin voluntarily surrenders lock-in when induced ecosystem investment is worth more than the additional developer rent, leakage, and transfer risk. But the entitlement is effective only when a destination recognizes it. Verification expands the privately optimal region when its safety and induced-investment gains exceed the outside-rent and leakage channels; otherwise the instruments can be substitutes. The social entitlement is higher when rents and diversion are mainly transfers and can be lower when migration creates substantial real duplication, fragmentation, or security cost.

The analysis recasts interoperability as a governance choice over investment. Communication standards determine whether agents can talk; export rights, destination recognition, and evidence determine whether developers will build capabilities worth communicating. Portability differs from a promise not to change fees: it protects exit while preserving adaptation. The next empirical step is to estimate capability spillovers, destination recognition, and investment responses on platform task graphs. The Online Appendix's bounded implementation audit is retained for reproducibility but is not part of the paper's economic evidence. The central conclusion is institutional: an origin may profit from granting a verified exit right because that right protects the relational capability base the origin later monetizes.

# Online Appendix: Developer Investment in Agentic Platforms

**Chupeng Xie**

This Online Appendix provides the task-composition microfoundation, proofs, destination-recognition and coordination details, bargaining and heterogeneity extensions, numerical details, the complete bounded software-policy audit, and the reproducibility map. Equation and proposition numbers without an appendix prefix refer to the main paper.

## Appendix A. Task Composition and Ecosystem Capability

### A.1 Serial workflows

Let $Z$ be a finite set of workflow types. Workflow $z$ arrives at rate $d_z \geq 0$, has value $V_z > 0$, and requires the agent set $S_z$. Conditional on investments, component success events are independent in the baseline. Agent $i$ succeeds with probability $q_i(e_i)$, where $q_i' > 0$ and $q_i'' \leq 0$. Expected completed-task value is

$$K(e) = \sum_{z \in Z} d_z V_z \prod_{j \in S_z} q_j(e_j). \tag{A1}$$

Define $Q_{z,-ij}$ as the success probability of every component in workflow $z$ other than agents $i$ and $j$. For any two agents that co-occur in at least one workflow,

$$\frac{\partial^2 K}{\partial e_i \partial e_j} = \sum_{z : i, j \in S_z} d_z V_z q_i'(e_i) q_j'(e_j) Q_{z,-ij} > 0. \tag{A2}$$

Equation (A2) formalizes productive complementarity. It does not require more users to join when an agent invests. Holding task arrivals fixed, the investment improves the marginal output of another required agent. If task success is not conditionally independent, (A2) continues to hold whenever the joint success technology is supermodular in component reliability.

## A.2 From the task graph to the reduced form

Suppose agents are ex ante symmetric and each workflow draws $m_z$ components from a large pool. Let $q(e)$ be common reliability. Then

$$K(e)=\sum_z d_z V_z q(e)^{m_z}. \tag{A3}$$

The main text writes ecosystem value as $G(Ne)$. This does not assert that (A3) depends only on the sum in every task graph. It is a local aggregate representation around a symmetric allocation. Define $E=Ne$ and

$$G(E)=K(E/N)-N\acute{F}(E/N), \tag{A4}$$

where $\acute{F}$ removes the directly attributed component already included in $F$. The conditions $G'>0$ and $G''\le 0$ correspond to positive but diminishing ecosystem completion value over the region studied. The serial microfoundation can have locally increasing returns when reliability is initially low. Appendix A.4 shows that the main marginal condition remains valid without global concavity; strict concavity is needed only for uniqueness and the three-region characterization.

## A.3 Capability stock versus traffic externality

Consider a payment agent $i$ and a procurement agent $j$. With fixed arrivals $d$ of procurement requests, value is

$$K(e_i,e_j)=dVq_i(e_i)q_j(e_j). \tag{A5}$$

Investment by the payment agent raises $\partial K / \partial e_j$ even if it attracts no new user and receives no separate direct request. By contrast, an anchor tenant raises $d$ by attracting traffic. Both mechanisms can coexist, but they are empirically separable. A task-graph design measures whether an intervention on $i$ changes success conditional on a fixed request, whereas a traffic design measures whether request arrivals change.

### A.4 Nonconcave task composition

If $K$ is locally convex, platform profit can have multiple local maxima in portability. The first-order marginal decomposition in (13) remains exact. The global solution is

$$p^{*} \in \arg\max_{p \in [0,1)} \Pi(p, v). \tag{A6}$$

The boundary conditions in Proposition 2 are then sufficient only with quasi-concavity or a verified global comparison. The released numerical routine evaluates the full bounded grid rather than relying on a local optimizer, so it remains valid in this extension. Economically, nonconcavity can generate a coordination threshold: a small entitlement induces too little capability to make a composed workflow viable, while a larger entitlement unlocks a discrete task set.

## Appendix B. Proofs and Primitive Conditions

### B.1 Ex post fee

Let $Y_i(e)$ denote the operating value attributable to retaining developer $i$ inside the platform, including any continuation value that the platform loses if $i$ exits. The platform offers an operating transfer $t_i$ after observing investment. The developer stays if $t_i \geq O(e_i, p)$. Under a take-it-or-leave-it offer and the feasibility condition $Y_i \geq O$, the platform chooses

$$t_i^*(e, p) = O(e_i, p). \tag{B1}$$

An access fee applied to developer gross revenue, a revenue-share reset, and a routing policy are payoff equivalent if each leaves (B1). The baseline abstracts from legal restrictions on discriminatory fees and from private information at this stage. Bargaining is introduced in Appendix F.

**B.2 Proof of Proposition 1**

Developer $i$ solves

$$\max_{e_i \geq 0} \left\{ r(p) e_i - \frac{c}{2} e_i^2 \right\}. \tag{B2}$$

The objective is strictly concave. Its first-order condition is $r(p) - c e_i = 0$, so the unique solution is $e_i = e(p) = r(p)/c$. Differentiation gives $e'(p) = r'(p)/c > 0$.

The symmetric social surplus, excluding institutional costs that do not depend on $e$, is

$$W(e) = N F(e) + G(N e) - \frac{N c}{2} e^2. \tag{B3}$$

Because $F'' \leq 0$, $G'' \leq 0$, and $c > 0$, (B3) is strictly concave. Its derivative divided by $N$ is $F'(e) + G'(N e) - c e$. At the decentralized investment $c e = r(p)$, this derivative is $\Delta(e; p)$. If $\Delta > 0$, the strict-concavity maximizer lies to the right, so $e(p) < e^{FB}$. Finally,

$$F'(e) + G'(N e) - r(p) = [F'(e) - r(p)] + G'(N e), \tag{B4}$$

which gives the decomposition. If $F'(e) = r(p)$ and $G'(N e) > 0$, (B4) remains positive.

For the bargaining extension, the developer maximizes (12A) minus investment cost. Differentiation gives (12B). Subtracting (12B) from the first-best marginal condition yields (12C). Implicit differentiation gives

$$e_p = \frac{(1-\theta) r'(p)}{c - \theta[F''(e) + N G''(N e)]} > 0, \tag{B4A}$$

because the denominator is positive under concavity. For every $\theta < 1$, a positive difference between social marginal value and the outside return therefore implies both underinvestment and a positive investment response to portability. Q.E.D.

### B.3 Derivation of the platform marginal value

Write the reduced profit as

$$\Pi = N F(e(p)) + G(N e(p)) - N r(p) e(p) - Q(p, v) - k(v), \tag{B5}$$

where $Q = C + T + H(1-v)$. Differentiating with respect to $p$ gives

$$\Pi_p = N[F'(e) + G'(N e) - r] e' - N r' e - Q_p, \tag{B6}$$

which is (13).

### B.4 Proof of Proposition 2

Under strict concavity, $M(p, v) = \Pi_p(p, v)$ is strictly decreasing in $p$. The Karush–Kuhn–Tucker conditions for $0 \le p \le 1$ are necessary and sufficient. If $M(0, v) \le 0$, concavity implies $M(p, v) < 0$ for every $p > 0$, so $p^* = 0$. Conversely, a maximum at zero requires the right derivative to be nonpositive. If $M(0, v) > 0 > M(1, v)$, continuity and strict monotonicity give a unique $p^* \in (0, 1)$ with $M(p^*, v) = 0$. If $M(1, v) \ge 0$, profit is increasing up to the upper boundary and $p^* = 1$; the converse follows from the left derivative condition. Evaluating $M(0, v) > 0$ gives (14). Q.E.D.

### B.5 Primitive strict concavity

For the numerical and experimental economic environment, let $r(p) = r_0 + \mu p$, so $e_p = \mu / c$ and $e_{pp} = 0$. Then

$$\Pi_{pp} = N[F''(e) + NG''(Ne)]\left(\frac{\mu}{c}\right)^2 - \frac{2N\mu^2}{c} - C'' - T'' - (1-v)H''. \quad \text{(B7)}$$

If $F'' \leq 0$, $G'' \leq 0$, $C'', T'', H'' \geq 0$, and $\mu > 0$, (B7) is strictly negative. Thus the three-region result is not created by imposing a single-peaked numerical curve; it follows from standard diminishing value and convex governance costs.

With nonlinear concave $r$, additional terms are

$$N\Delta(e;p)e''(p) - Nr''(p)e(p). \quad \text{(B8)}$$

Since $e'' = r''/c \leq 0$, a sufficient condition is $\Delta \geq ce$, together with the curvature conditions above. More generally Assumption 2 can be checked directly from primitives.

## B.6 Comparative statics in ecosystem value

Let $G(E;\beta)$ satisfy $G'_\beta(E) > 0$. At an interior solution, implicit differentiation of $M(p^*, v;\beta) = 0$ yields

$$\frac{\partial p^*}{\partial \beta} = -\frac{NG'_\beta(Ne)e'(p)}{\Pi_{pp}} > 0. \quad \text{(B9)}$$

The result also holds as monotone movement of the optimal correspondence when $G$ has increasing differences in $(E, \beta)$.

## B.7 Proof of Proposition 3

Differentiate destination payoff (15). Because $v$ enters only through expected import harm,

$$\Pi^B_{qv} = pL'(pq) \geq 0. \quad \text{(B10)}$$

Strict concavity makes the recognition choice unique. At an interior solution,

$$q^*_v(p, v) = -\frac{\Pi^B_{qv}}{\Pi^B_{qq}} \geq 0, \quad \text{(B11)}$$

and the boundary result follows from monotone comparative statics.

For the origin, differentiate (17). First,

$$\Pi_p^A = N\,\Delta e_p - N\,r_p e - Q_p. \tag{B12}$$

Differentiating once more with respect to $v$ gives

$$\Pi_{pv}^A = N\left[\left(s'' e_v - r_v\right) e_p + \Delta e_{pv} - r_{pv} e - r_p e_v\right) - Q_{pv}. \tag{B13}$$

Developer optimization implies $r_v = c\,e_v$, so (B13) is exactly (18). Increasing differences are equivalent to $\Psi \geq 0$; strict negativity gives local strategic substitutability. Moving all non-safety terms to the right yields sufficient condition (18A). If recognition is fixed, $r_v = e_v = 0$. If the only $p$–$v$ interaction in cost is $H(p)(1-v)$, then $Q_{pv} = -H'(p)$ and (18) reduces to $\Pi_{pv}^A = H'(p) \geq 0$. Q.E.D.

**B.8 Proof of Proposition 4**

Substitute $x = z - \mu\,p\,q$ into (21). Strict convexity of $J$ and $Q$ makes the reduced objective strictly convex and the minimizer unique. The Lagrangian is

$$L = J(x) + Q(p, v; q) + \lambda(z - x - \mu\,p\,q). \tag{B14}$$

The interior first-order conditions are $J'(x) = \lambda$ and $Q_p = \lambda\,\mu\,q$, which combine to give (22). The KKT inequalities imply a portability corner when $Q_p/(\mu\,q)$ lies below the corresponding commitment marginal cost throughout the feasible set, and a commitment corner under the reverse ordering. A pure portability implementation of target $z \leq \mu\,q$ uses $p = z/(\mu\,q)$ and costs $Q(z/(\mu\,q), v; q)$; a pure commitment implementation uses $x = z$ and costs $J(z)$. Their strict comparison gives (22A). Q.E.D.

### B.9 Proof of Proposition 5

Differentiating (23) gives

$$W_p = N[F'(e) + G'(Ne) - ce]e' - C' - (1-v)H' - \omega T' - D'. \tag{B15}$$

Because developer optimization implies $ce = r(p)$, subtracting (13) from (B15) gives (24). If $\Omega \geq 0$ everywhere and $p^P > p^S$, concavity of welfare implies $W_p(p^P) \leq 0$, while private optimality and (24) imply $W_p(p^P) \geq \Pi_p(p^P) = 0$ for an interior private optimum. The strict case yields a contradiction; boundary cases follow from the KKT inequalities. Hence $p^P \leq p^S$. Applying the same argument with $\Omega \leq 0$ and interchanging the objectives gives $p^P \geq p^S$. If $\Omega$ changes sign, either ordering can be constructed while preserving concavity by shifting the curvature and level of $D$, so no global order follows. Q.E.D.

## Appendix C. Subsidies and Heterogeneous Spillovers

### C.1 Investment subsidy

Suppose the platform pays a linear capability subsidy $s_i e_i$ before its ex post fee choice. If the subsidy is legally protected from clawback, developer investment becomes

$$e_i = \frac{r_i(p_i) + s_i}{c_i}. \tag{C1}$$

The subsidy is a direct payment and portability is an outside-option right. Holding their induced marginal return equal, a subsidy costs the platform $s_i e_i$ regardless of exit, while portability costs additional retained return only through the ex post participation constraint and adds governance costs. If the platform can perfectly measure developer $i$'s workflow spillover, it can target $s_i$ to that contribution. If spillover is hard to measure but ex post appropriation is severe, a common portability entitlement can protect decentralized information about valuable investments.

The instruments need not be substitutes. Suppose a developer has a high observable spillover but still expects the platform to appropriate any private complement to the subsidized project. A contribution subsidy addresses the $G'$ wedge; portability addresses the $F'-r$ wedge. Both are required to implement first best. If instead a complete subsidy contract conditions payment on all direct and ecosystem output and is immune to renegotiation, portability is redundant for investment incentives, though it may retain competition benefits.

## C.2 Heterogeneous developers

Let total gross value be $Y(e)$, with $Y_i > 0$ and a negative-semidefinite Hessian. Developer $i$ has cost $c_i e_i^2/2$ and retained return $r_i(p_i)e_i$. Then

$$e_i(p_i) = \frac{r_i(p_i)}{c_i}. \tag{C2}$$

The platform's marginal value of developer-specific portability is

$$\Pi_{p_i} = \left[Y_i(e) - r_i(p_i)\right] e_i'(p_i) - r_i'(p_i) e_i - Q_{i,p_i}. \tag{C3}$$

The term $Y_i$ includes direct value and all workflows in which $i$ participates. Portability is therefore stronger for agents with larger task-graph centrality, greater investment response, lower conversion cost, and safer credentials. This is a contribution rule, not a popularity rule. A stand-alone high-volume agent can have lower $Y_i - F_i'$ than a low-volume permissions agent used by many composed workflows.

### C.3 Common standard and selective subsidy

If the platform must choose a common $p$, it averages (C3) across developers. A common standard can be below the individually optimal entitlement for high-spillover agents and above it for low-spillover agents. Selective subsidies can restore targeting without making the export schema discretionary. This division has a commitment advantage: a nondiscriminatory portability standard is difficult to withdraw from a successful developer, while a transparent contribution subsidy can vary with estimated spillovers.

## Appendix D. Destination Recognition and Coordination

### D.1 Recognition primitives

Main-text destination payoff (15) uses recognized imported capability

$$m(p,q) = pq\frac{r_0 + \mu pq}{c}. \tag{D1}$$

Its recognition derivatives are

$$m_q = \frac{p}{c}(r_0 + 2\mu pq) > 0, m_{qq} = \frac{2\mu p^2}{c} > 0. \tag{D2}$$

The convexity of imported quantity means strict concavity of $D$ alone is not sufficient for a unique recognition decision. A primitive sufficient condition is

$$D''(m)m_q^2 + D'(m)m_{qq} - K_B''(q) - (1-v)p^2 L''(pq) < 0. \tag{D3}$$

Under (D3), the first-order condition in (16) is sufficient. Verification raises recognition because $\Pi_{qv}^B = pL'(pq) \geq 0$. Entitlement also raises recognition whenever

$$\Pi_{qp}^B = D''(m)m_p m_q + D'(m)m_{pq} - (1-v)\left[L'(pq) + pqL''(pq)\right] \geq 0. \tag{D4}$$

Condition (D4) says that the destination's productive value from a larger eligible import must dominate the additional import-risk margin. When it holds, $q^*(p,v)$ is nondecreasing in both the origin's entitlement and evidence reliability.

### D.2 A low-portability/low-recognition trap

Let the origin's incremental return from a recognized import be $A(pq)$ and the destination's be $B(pq)$, with $A(0)=B(0)=0$. Suppose implementing a positive export entitlement and a positive recognition rule requires fixed costs $\kappa_A$ and $\kappa_B$. Reduced governance payoffs are

$$V_A(p,q) = A(pq) - \kappa_A 1\{p>0\} - \frac{c_A}{2}p^2, V_B(p,q) = B(pq) - \kappa_B 1\{q>0\} - \frac{c_B}{2}q^2. \tag{D5}$$

At $q=0$, no positive $p$ generates recognized value, and at $p=0$, no positive $q$ generates an eligible import. Hence $(0,0)$ is a Nash equilibrium whenever fixed costs are positive. If there exists $(\hat{p},\hat{q}) \gg 0$ such that $V_A(\hat{p},\hat{q})>0$ and $V_B(\hat{p},\hat{q})>0$, the zero equilibrium is Pareto dominated by coordinated adoption. A reciprocal-recognition agreement can select the positive outcome; a standard-setting organization can do so by reducing $\kappa_A$ and $\kappa_B$. Neither intervention implies that $p=q=1$ is efficient.

This argument proves Corollary 1. Conditional adoption removes the unilateral fixed-cost loss: each platform incurs its integration cost only if the other adopts the agreed positive level. Participation is individually rational at $(\hat{p}, \hat{q})$ by the two strict inequalities, while the agreement leaves the zero outcome available if either party declines.

### D.3 Recognition without commitment

The main text treats destination recognition as a public, auditable rule announced before investment. To expose the role of that commitment, suppose instead that $B$ announces $q_0$ but, after developer investment, can reset recognition to $q_1$. Let $B$ obtain contemporaneous import value $D(m)$ but pay per-unit liability or integration cost $\ell q_1$, while the developer's continuation value remains $r_0 + \mu p q_1$. If the installed capability does not enter $B$'s current payoff except through the imported record, the ex-post choice solves

$$q_1^*(p, e) \in \arg\max_{q \in [0,1]} \{D(pqe) - K_B(q) - (1-\nu)L(pq) - \ell q\}. \tag{D6}$$

The developer anticipates $q_1^*$ rather than $q_0$ and invests from the protected return $r_0 + \mu p q_1^*$. Any ex-post state in which $q_1^* < q_0$ therefore creates a destination-side hold-up wedge in addition to the origin-side fee wedge. A reciprocal-recognition contract is valuable precisely when it makes $q_0$ verifiable and enforceable before investment. If $B$'s recognition rule is noncontractible, all baseline investment formulas remain valid after replacing $q$ by the anticipated $q_1^*$, but effective portability and investment are weakly lower whenever the reset reduces recognition. Thus the baseline commitment assumption is a benchmark for an auditable protocol rule, not a claim that every destination can bind every future policy.

## Appendix E. Fee Commitment and Portability

Let $x \geq 0$ be a protected marginal return supplied by a fee or routing commitment. Holding destination recognition $q > 0$ fixed at its equilibrium value, total retained marginal return is

$$R(x, p; q) = r_0 + x + \mu p q, \quad e(x, p) = \frac{R(x, p)}{c}. \tag{E1}$$

Let $J(x)$ be the enforcement and rigidity cost of commitment, and let $Q(p, v) = C + T + H(1 - v)$. Platform profit is

$$\Pi(x, p, v) = N F(e) + G(N e) - N R e - J(x) - Q(p, v) - k(v). \tag{E2}$$

At an interior solution, define $\Delta = F'(e) + G'(N e) - R$. The two protection conditions are

$$\Pi_x = \frac{N \Delta}{c} - N e - J'(x) = 0, \tag{E3}$$

and

$$\Pi_p = \mu q \left[ \frac{N \Delta}{c} - N e \right) - Q_p(p, v; q) = 0. \tag{E4}$$

Combining (E3) and (E4) yields

$$J'(x^*) = \frac{Q_p(p^*, v; q)}{\mu q}. \tag{E5}$$

The platform equates the marginal institutional cost per unit of protected return. Higher fee-contract rigidity shifts protection toward portability. Higher migration, privacy, and stale-record cost shifts protection toward commitment. Verification shifts the mix toward portability only when it lowers $Q_p$ enough to dominate any recognition-induced leakage. This is the same conditional logic as Proposition 3.

Commitment and portability differ beyond this static cost comparison. A fee promise must describe future contingencies. Portability can remain valid while the platform changes price, removes an unsafe integration, or redesigns routing. A narrow, enforceable export right can therefore dominate a broad promise not to adapt.

## Appendix F. Ex Post Bargaining

The take-it-or-leave-it benchmark gives the platform all incremental surplus. Let developer bargaining weight be $\theta \in [0,1]$. Define per-developer inside value under symmetry as

$$y(e) = F(e) + \frac{G(Ne)}{N}. \tag{F1}$$

A Nash split gives the developer operating return

$$u^{op}(e,p) = (1-\theta)\, r(p)\, e + \theta\, y(e). \tag{F2}$$

The investment condition is

$$(1-\theta)\, r(p) + \theta\left[F'(e) + G'(Ne)\right] - c\,e = 0. \tag{F3}$$

Comparing (F3) with the first-best condition shows that the remaining wedge is

$$F'(e) + G'(Ne) - c\,e = (1-\theta)\left[F'(e) + G'(Ne) - r(p)\right]. \tag{F4}$$

Bargaining reduces but does not eliminate underinvestment for $\theta < 1$. Implicit differentiation gives

$$e_p = \frac{(1-\theta)\, r'(p)}{c - \theta\left[F''(e) + N\,G''(Ne)\right]} > 0. \tag{F5}$$

Hence portability continues to protect investment whenever the platform retains some bargaining power. As $\theta \to 1$, investment approaches first best and the investment rationale for portability disappears; competition and user-control rationales may remain.

## Appendix G. Platform Competition and Multihoming

The main text gives the destination an endogenous recognition decision but does not model symmetric price competition between origins. Let destination competition raise the value paid for recognized capability to $\acute{r}\, p\, q\, e$. Stronger competition raises the developer's investment response, but it also raises the origin's rent and leakage costs. The sign of the origin's portability choice remains governed by (13) after replacing $r_p$ with its total derivative.

If developers multihome before investing, some code and task experience are already portable, raising $r_0$. This reduces the hold-up wedge and the marginal investment benefit of a formal entitlement. A platform may therefore offer less additional portability in a market where multihoming is technically easy. That observation does not imply that formal rights are unimportant: user-authorized memory, platform-issued reputation, and routing records can remain nonportable even when code runs on several platforms.

Competing platforms may also underprovide a common verification standard. Each benefits when imported claims are reliable but may prefer that rivals pay the fixed schema and certification cost. A standard-setting organization can lower fixed evidence and recognition costs. Proposition 3 predicts greater portability only when the resulting safety and induced-investment channels dominate the added outside rent and competitive leakage; Appendix D.2 separately shows how a shared standard can eliminate a Pareto-dominated zero-adoption equilibrium.

# Appendix H. Numerical Implementation

## H.1 Functional form

The numerical illustration uses

$$F(e) = a e - \frac{b}{2} e^2, G(N e) = \beta \log(1 + N e), \tag{H1}$$

$$r(p) = r_0 + \mu p, T(p) = \frac{\ell}{2} p^2, H(p) = \frac{h}{2} p^2, k(v) = \frac{\kappa}{2} v^2. \tag{H2}$$

The baseline parameter vector is

| Parameter | Value | Role |
|---|---|---|
| $N$ | 6 | third-party agents |
| $a$ | 1.50 | direct marginal capability value |
| $b$ | 0.80 | direct-value curvature |
| $c$ | 4.00 | investment-cost curvature |
| $r_0$ | 0.80 | closed-platform outside return |
| $\mu$ | 2.00 | portability protection strength |
| $\beta$ | 8.00 | ecosystem spillover value |
| $\ell$ | 0.80 | competitive leakage curvature |
| $h$ | 3.00 | unverifiable-transfer harm |
| $\kappa$ | 0.80 | verification-cost curvature |

## H.2 Verification checks

The script computes $p^*$ on a 10,001-point grid and the regime map on a 2,001-point grid for each parameter cell. It checks the analytical derivative (13) against a centered finite difference with step $10^{-5}$. The maximum absolute error is reported in numerical_verification.json. It also asserts that the numerical domain contains at least one closed, partial, and full-portability cell.

The joint baseline optimum and first-best capability investment are exported in numerical_verification.json. These values describe the chosen normalization, not a field calibration.

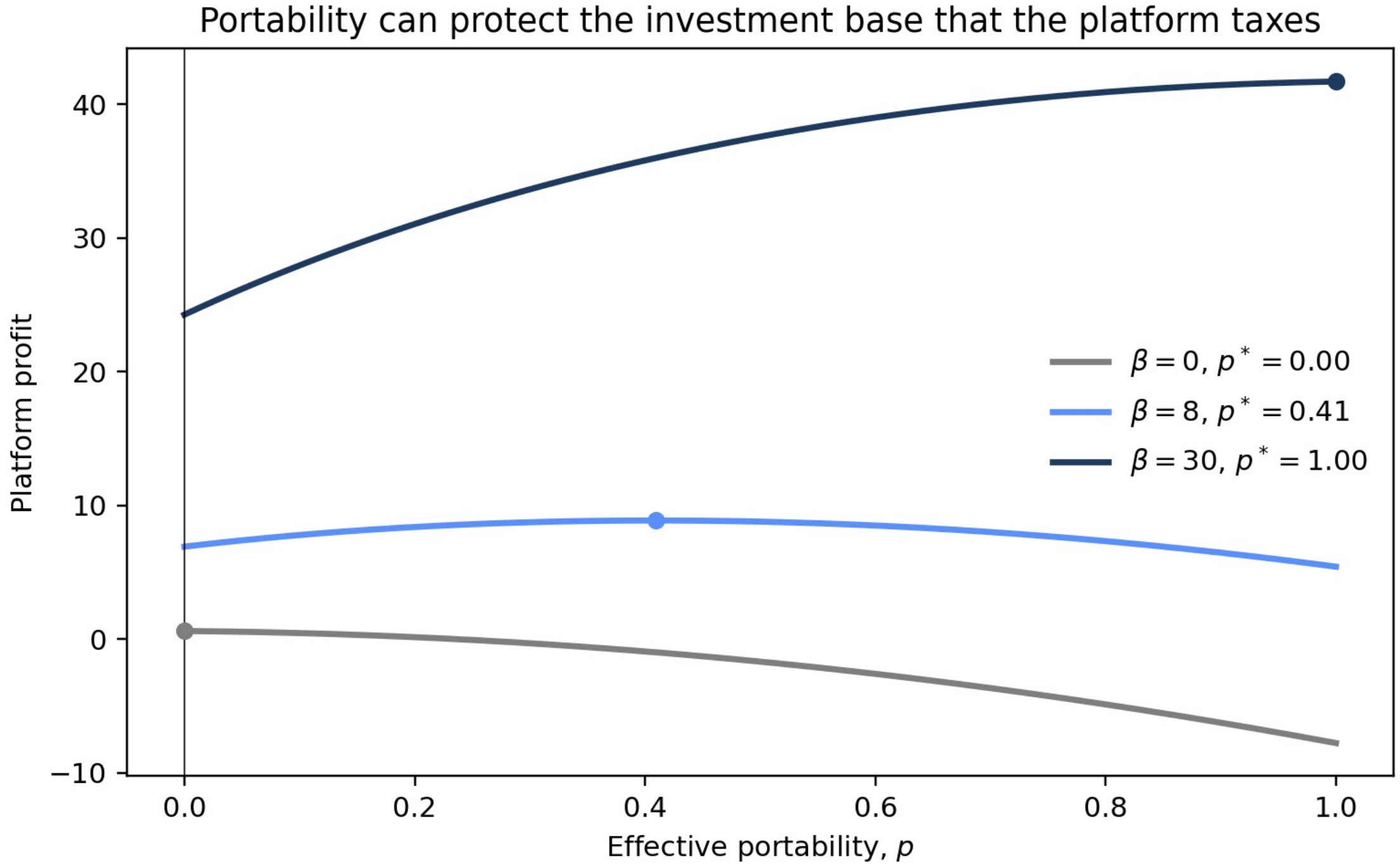

*Figure OA.1. Platform profit across portability levels.*

Figure OA.1 holds verification fixed and plots profit for three spillover values. With no ecosystem spillover, the direct hold-up channel supports little protection. At an intermediate value, profit is single-peaked at a partial entitlement. With a large spillover, the peak moves toward full portability. The curves provide numerical witnesses for all three regions of Proposition 2.

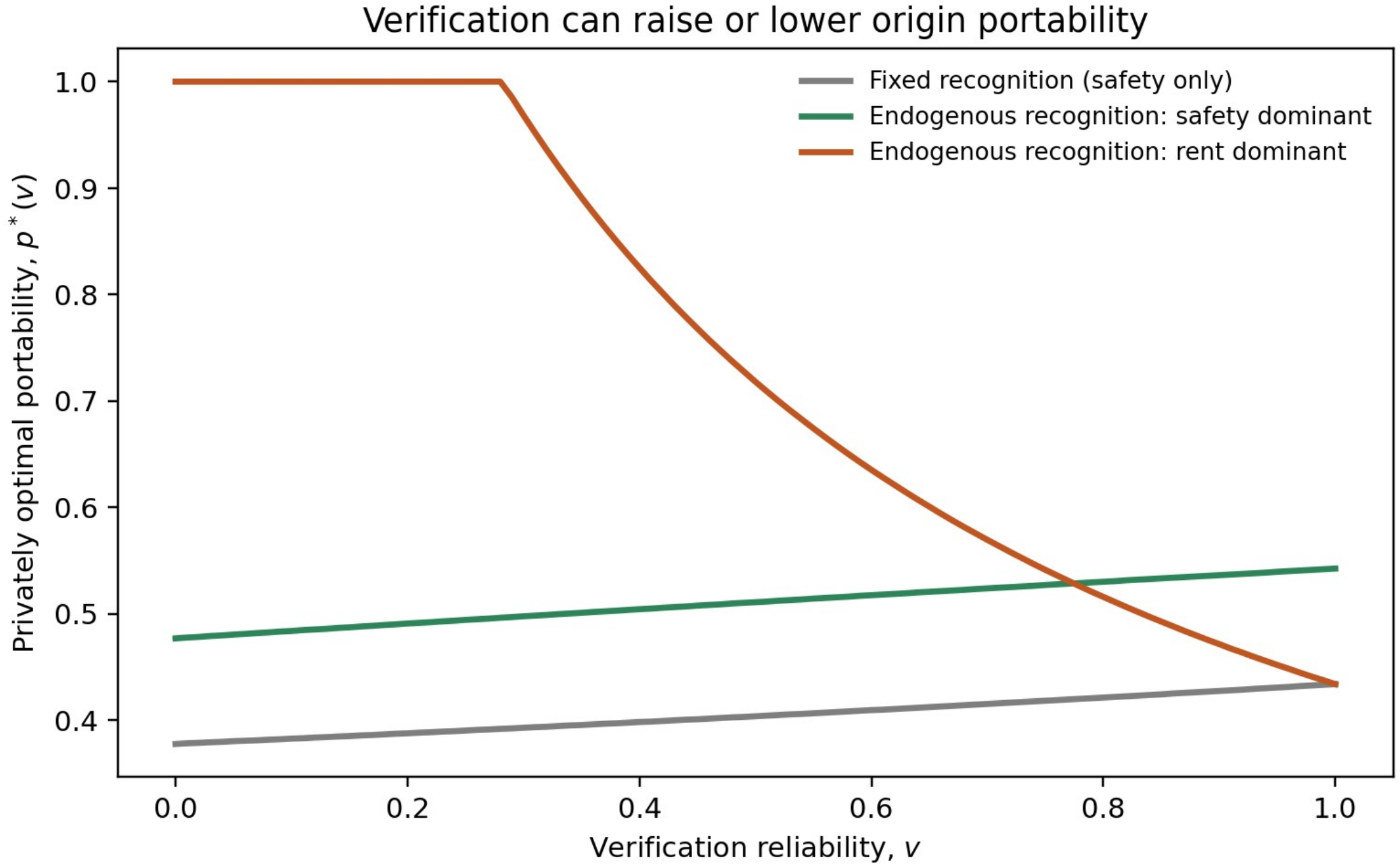

*Figure OA.2. Verification can raise or lower origin portability.*

Figure OA.2 adds two recognition-sensitive witnesses. In the safety-dominant case, destination recognition is $q(v)=0.70+0.10\,v$ and false-transfer curvature is 3.00; optimal origin portability rises with verification. In the rent-dominant case, $q(v)=0.20+0.80\,v$ and false-transfer curvature is 0.10; optimal origin portability falls. The code asserts both directions. The comparison holds the remaining primitives fixed and exists to demonstrate that Proposition 3 has nonempty complement and substitute regions.

### H.3 Private and social optima

The baseline welfare code excludes developer retained return and competitive leakage because they are transfers, but retains investment cost, verification cost, and unverifiable-transfer harm. This implements the $\omega=0, D=0$ region of Proposition 5. Users can reclassify a fraction of leakage as a real resource loss and add duplication or fragmentation cost $D$. As those real costs rise, the social optimum moves toward the private optimum and can fall below it.

# Appendix I. Bounded Software-Policy Implementation Audit

## I.1 Freeze and provenance

The confirmatory plan and complete task library were committed and pushed before outcome collection. The collector stores SHA-256 hashes of tasks.json, run.py, analyze.py, and preregistration.md; the output of ollama show for every model; the operating system and Python version; timestamps; prompts; raw model responses; parsed actions; retry count; latency; oracle action; and regret.

The freeze commit is 0f4f249. Smoke outputs use a separate run identifier and are excluded from confirmatory analysis. A final run cannot resume if any frozen hash or model digest differs. The audit trail also records two technical corrections made before any aggregate hypothesis test: commit 7e72cf4 replaces an unintended global model shuffle with the prespecified within-model task randomization, and commit c1ae4fb preserves the original analysis while adding a script that implements the preregistration's stated 12-scenario clustering rule. The first interrupted run and the original analysis script remain available; neither contributes to reported estimates. The amendment records the exact observation counts and timing.

## I.2 Study A task generation

Twelve environments were generated with seed 20260715. Baseline retained marginal return is drawn from $\{2.4, 2.8, 3.2, 3.6, 4.0\}$, the protection increment from $\{1.6, 2.0, 2.4, 2.8\}$, and cost curvature from $\{1.2, 1.4, 1.6, 1.8, 2.0\}$. The allowed investments are the integers zero through six. Every environment has an oracle investment increase under both protected institutions. Fee commitment and portability use the same numerical increment.

The system prompt identifies the model as a third-party developer policy in a simulated research environment and instructs it to maximize numerical net payoff. The user prompt states the institution, $r$, $c$, the payoff function, and the action set. It does not provide the payoff table or oracle.

### I.3 Study B task generation

Twelve environments vary capability spillover over $\{0,1,3,5,8,11,14,18,22,26,30,34\}$. Direct value, leakage, and risk curvature are frozen draws. For each candidate portability level, code produces gross revenue, developer retained return, competitive leakage, and portability risk from the model in Appendix H. The prompt reports these four components but not their net. The software policy must subtract costs and choose the maximum.

Low and high verification are 0.20 and 0.80. The frozen oracle is weakly higher under high verification in every environment and strictly higher in five of twelve, given the discrete action grid. The task library includes closed, partial, and full oracle choices.

### I.4 Hypotheses and analysis

H1a is the paired fee-commitment minus closed-platform investment difference. H1b is the paired portability minus closed-platform difference. H2 is the high-minus-low verification portability difference. Pairing is within model, task, and repetition. H3 is the scenario-level Pearson correlation between spillover and mean chosen portability, with high verification primary.

The analysis reports scenario-cluster bootstrap intervals and cluster sign-flip randomization p-values. Model-specific results, invalid-output rates, oracle accuracy, policy error, regret, and latency are descriptive. Generated explanations are retained for audit but never coded as confirmatory outcomes.

### I.5 Complete results

The collector completed all 600 planned calls. Twenty-four calls were invalid after the allowed retry, leaving 576 valid actions. Table I1 reports the analysis that implements the preregistration's stated task-scenario clustering. H1a and H1b use 100 valid paired model-task-repetition contrasts; missing pairs arise only from retained invalid calls. These pairs are averaged within each of the 12 frozen tasks before bootstrap and sign-flip inference. H2 has all 120 pairs. H3 is defined directly at the 12-scenario level.

**Table I1. Confirmatory institutional comparisons**

| Comparison | Scenarios | Raw pairs | Estimate | 95% scenario-bootstrap interval | Sign-flip $p$ |
|---|---|---|---|---|---|
| H1a: fee commitment minus closed investment | 12 | 100 | -0.0486 | [-0.1653, 0.0833] | 0.5341 |
| H1b: portability minus closed investment | 12 | 100 | -0.0528 | [-0.1708, 0.0833] | 0.4671 |
| H2: high minus low verification portability | 12 | 120 | 0.0500 | [0.0292, 0.0750] | 0.0077 |
| H3: spillover–portability correlation, high verification | 12 | – | 0.6563 | [0.1930, 0.8758] | – |
| H3: spillover–portability correlation, low verification | 12 | – | 0.6385 | [0.1653, 0.8845] | – |

The first two estimates reject neither zero nor modest effects in either direction. This is not an absence created by the task library: the code oracle raises investment by 1.33 units on average under both fee commitment and portability. Rather, several artifacts used nearly invariant choices. The governance tasks show a different pattern. Higher verification moves chosen portability upward, and scenarios with greater capability-spillover value receive greater portability under both verification conditions. The directional results coexist with poor level calibration.

**Table I2. Artifact-level execution performance**

| Model artifact | Calls | Invalid rate | Oracle-optimal rate | Mean policy error | Mean regret | Mean seconds |
|---|---|---|---|---|---|---|
| gemma3:4b | 120 | 0.000 | 0.250 | 0.638 | 2.032 | 11.402 |
| llama3.1:8b | 120 | 0.017 | 0.390 | 0.564 | 1.549 | 8.887 |
| qwen3:4b | 120 | 0.150 | 0.098 | 2.015 | 5.638 | 3.482 |
| qwen3:8b | 120 | 0.033 | 0.052 | 1.047 | 2.278 | 3.868 |
| qwen3:14b | 120 | 0.000 | 0.333 | 0.821 | 2.081 | 6.120 |
| Overall | 600 | 0.040 | 0.229 | – | 2.631 | – |

The results identify an execution boundary, not a contradiction in the economic model. The propositions describe optimal organizational choices given the stated objective. The benchmark asks whether a language-model policy computes those choices from a prompt. Directional movement in portability can occur even when only 22.9% of level choices are optimal. Conversely, failure to raise investment under a protected return shows that an organization cannot infer reliable implementation from a model's fluent explanation or valid JSON alone. A deterministic optimizer, payoff checker, or approval control remains necessary for these tasks.

For transparency, the originally frozen analysis script is retained. It incorrectly passed all model-by-repetition pairs directly to procedures labeled as scenario-clustered. Its point estimates are nearly the same but its nominal cluster counts are 100 or 120 rather than 12. The correction was documented after 29 of 600 replacement-run decisions and before any aggregate contrast or hypothesis test. Table I1 uses the clustering rule stated in the preregistration; both outputs and the amendment are released.

### I.6 Limits

The study has five artifacts but only three model families, all run locally through one inference engine. Temperature-zero output can still vary across hardware or software versions; recorded hashes and seeds reduce but do not eliminate that risk. The economic environments are stylized and fully disclosed to the model. The benchmark therefore tests arithmetic and institutional policy execution, not learning from experience, strategic deception, or organizational adoption. A field study would require actual platform policy variation, developer investment, task-graph outcomes, and legally authorized data.

## Appendix J. Reproducibility Map

| Claim or artifact | Source |
|---|---|
| Propositions 1–5 | manuscript.md; proofs in Appendix B |
| Figures 1–4 | numerics.py |
| Figure source data | figures/regime_map.csv;<br>figures/verification_comparative_statics.csv |
| Numerical checks | figures/numerical_verification.json |
| Frozen hypotheses | experiment/preregistration.md |
| Frozen task generator | experiment/make_tasks.py |
| Frozen task library | experiment/tasks.json |
| Model collector | experiment/run.py |
| Confirmatory analysis | experiment/analyze.py |
| Raw and tabular outcomes | experiment/results/ |
| Word/PDF builder | ../build.py |

From the papers/ai-agent-markets/investing-agentic-platforms directory, run:

```
python3 numerics.py
cd experiment
python3 run.py --resume
python3 analyze.py results/agentic-platform-governance-v1.csv --output results/analysis.json
cd ../..
python3 build.py --paper platform-investment
python3 build.py --paper platform-investment-anonymous --skip-numerics
python3 build.py --paper platform-investment-appendix --skip-numerics
python3 build.py --paper platform-investment-appendix-anonymous --skip-numerics
```

The collector refuses to overwrite an existing run without --resume and refuses a resume if frozen metadata differ. This protects the confirmatory dataset from silent prompt, task, model, or analysis changes.

## Appendix K. Field Identification Agenda

The paper's model can be taken to platform data without treating the open-weight benchmark as a substitute for field behavior. The required unit is a task graph, not merely an agent listing. For each task, the platform would need the task type, agents invoked, versions, routing decision, completion outcome, gross value, compute and verification cost, and any human intervention. For each developer, it would need dated investments in connectors, tests, workflows, and portable records. Governance data would include fee schedules, routing rules, export rights, verification requirements, and policy changes.

### K.1 Identifying productive capability spillovers

The key parameter is the effect of agent $i$'s investment on workflows that use other agents. A clean experiment randomizes access to a new connector or verified version across otherwise eligible workflow clusters. Let $Z_{gt}$ indicate that cluster $g$ receives the integration at time $t$, let $Y_{jgt}$ be completion value for workflows routed to complement $j \neq i$, and let $C_{ijg}$ indicate that agents $i$ and $j$ co-occur in the cluster's workflows. A reduced-form design is

$$Y_{jgt} = \alpha_g + \lambda_t + \beta_1 Z_{gt} + \beta_2 Z_{gt} C_{ijg} + \varepsilon_{jgt}. \tag{K1}$$

The second interaction distinguishes productive composition from a platform-wide demand shock. The experiment should hold task arrivals or condition on a common arrival pool. If the integration also attracts new tasks, arrivals and conditional completion must be reported separately.

A staged rollout can substitute for randomization when timing is plausibly unrelated to local task shocks. Event-study pretrends are necessary but not sufficient because platforms may prioritize clusters expected to grow. Instrumenting deployment with engineering capacity, compatibility batches, or exogenous security-review queues can improve identification if the exclusion restriction is credible.

### K.2 Identifying hold-up

Fee increases alone do not identify hold-up; they may reflect cost. The prediction concerns investment made before a discretionary policy change. A platform can randomize the duration of a fee or routing guarantee among newly admitted developers. Investment is then compared before any ex post change occurs. The primary outcomes are connector completion, evaluation coverage, workflow depth, and safety testing—not only entry or self-reported confidence.

An observational design can use unexpected changes in revenue share, routing, tool price, or first-party priority. Let exposure depend on the fraction of developer value tied to the changed platform resource before the announcement. A difference-in-differences design compares high- and low-exposure developers, with investment measured from code, testing, and integration traces. The identifying assumption is that absent the policy change, investment trends would have been parallel. Strategic anticipation is a direct threat; pre-announcement investment declines should be estimated rather than discarded.

### K.3 Identifying the portability response

A portability pilot should vary effective outside value, not merely offer a download button. Treatment must include a destination-readable schema, authorized export, and recognition of the record by at least one destination. Outcomes include investment before any migration, migration exercise, retained task success after migration, and false-transfer incidents. This separates $r'(p)$ from $H(p)(1-v)$.

A factorial pilot can randomize portability entitlement and evidence reliability while separately measuring destination recognition. The model predicts that recognized entitlement raises investment. Verification raises the origin's willingness to offer entitlement only when the measured reduction in transfer harm and induced-investment value exceeds the increase in outside rent and leakage. If developers, the origin, and the destination choose at different stages, all three decisions should be measured separately. A developer response to a "verified" label is not evidence for Proposition 3, whose decisions are recognition by the destination and entitlement by the origin.

### K.4 Structural estimation

With repeated policy variation, the platform problem can be estimated structurally. Developer investment first identifies the retained-return schedule from

$$c_i'(e_i) = r_i(p, q, v, x), \tag{K2}$$

where $x$ captures fee protection. Task outcomes and routing identify direct and ecosystem value. Export incidents and destination recognition identify real conversion cost and verification-sensitive harm. The platform's observed policy choice then disciplines competitive leakage and any unobserved governance cost. Counterfactuals compare closed, partially portable, fully portable, fee-committed, and subsidized regimes.

The strongest design combines randomized institutional variation with structural estimation. Purely revealed policy is selected by the platform and cannot identify the profit function without instruments or functional restrictions. Conversely, a small experiment may not cover the long-run state space. The model supplies cross-equation restrictions while the experiment supplies exogenous movement.

### K.5 Falsifiable predictions

The mechanism would be weakened by any of the following findings:

1. effective portability does not increase pre-migration capability investment;
2. investments with high workflow centrality receive no larger platform benefit than stand-alone investments;
3. verification does not reduce false, stale, or unauthorized transfer harm;

4. platforms with greater captured ecosystem value are not more willing to protect developer exit;
5. fee commitment raises investment but an economically equivalent outside option does not.

The last comparison is particularly informative. If only a fee promise changes investment, portability may not create a credible or usable outside return. If both work, the investment-protection interpretation is supported. If portability works more strongly, it may carry competition or user-control benefits beyond the baseline.

## Appendix L. Mapping Verification Architectures to the Model

The model does not select a named technology. Architectures differ in the cost of producing evidence, the claims they can support, and the residual harm after verification. Table L1 maps common implementations to primitives.

| Architecture | Evidence strength | Main residual risk | Model effect |
|---|---|---|---|
| Ordinary disclosure | developer statement | strategic misreporting, no execution binding | low $v$, low direct cost |
| Independent audit | sampled process and records | sampling error, delay, auditor incentives | higher $v$, recurrent $k(v)$ |
| Signed execution log | identity, time, declared inputs and outputs | garbage-in, key compromise, omitted off-log action | higher provenance component of $v$ |
| Trusted execution environment | code and state inside measured hardware boundary | side channels, hardware trust, unmeasured external tools | lowers execution component of $H$ |
| Zero-knowledge attestation | compliance with a formal predicate without revealing inputs | incomplete predicate, setup and proving cost | high rule-verification $v$, potentially high $k$ |
| Append-only or distributed ledger | ordering, persistence, attribution of submitted records | truth of submitted claims, privacy, governance | lowers tampering risk; does not alone establish capability |

### L.1 Evidence coverage

Let verification have components $v = (v_I, v_A, v_E, v_O)$ for identity, authorization, execution, and outcome. Transfer harm can be written

$$H(p,v) = H_I(p)(1 - v_I) + H_A(p)(1 - v_A) + H_E(p)(1 - v_E) + H_O(p)(1 - v_O). \tag{L1}$$

The platform chooses coverage by comparing component-specific harm with cost. A signed log may make attribution nearly certain while leaving outcome quality unknown. A zero-knowledge proof can establish that a price or permission rule was followed while saying nothing about whether the underlying task estimate was accurate. “Verified agent” is therefore not a sufficient economic category.

### L.2 Privacy and selective disclosure

Portability can increase privacy risk if raw task traces reveal principals, counterparties, or proprietary tools. Selective disclosure changes the cost function. A proof that a test score exceeds a threshold can support destination recognition without exporting the test data. The relevant comparison is not cryptography versus no cryptography; it is whether the reduction in $H$ justifies the proving, integration, governance, and revocation cost in $k$.

### L.3 Control-layer boundary

An append-only record is the evidentiary backbone of a control layer, not the complete control layer. A production system also needs authorization checks before action, policy evaluation, exception handling, rollback where feasible, and allocation of liability when rollback is impossible. These functions can reference a common record while remaining implemented through databases, gateways, trusted hardware, or human approval. The paper’s portability result concerns which authenticated state should remain usable after exit; it does not imply that every control decision should move on-chain.